\documentclass[useAMS,usenatbib]{mn2e}

\usepackage{amssymb,amsmath}
\usepackage{longtable}
\usepackage{dcolumn}
\usepackage{graphicx}

\usepackage{mathrsfs}

\usepackage{natbib}
\title[A new method to measure the virial factors]{A new method to measure the virial factors in the reverberation mapping of AGNs}

\author[H. T. Liu, H. C. Feng and J. M. Bai]
{H. T. Liu$^{1,3}$\thanks{E-mail: htliu@ynao.ac.cn}, H. C.
Feng$^{1,2,3}$ and J. M. Bai$^{1,3}$ \\
$^{1}$Yunnan Observatories, Chinese Academy of
Sciences, Kunming, Yunnan 650011, China\\
$^{2}$University of Chinese Academy of Sciences,
Beijing 100049, China\\
$^{3}$Key Laboratory for the Structure and Evolution of Celestial
Objects, Chinese Academy of Sciences, Kunming, Yunnan 650011,
China}

\begin{document}

\date{Accepted . Received}


\maketitle

\label{firstpage}

\begin{abstract}
Based on the gravitational redshift, one prediction of Einstein's general relativity theory, of broad optical emission lines in active
galactic nuclei (AGNs), a new method is proposed to estimate the virial factors $f$ in measuring black hole masses $M_{\rm{RM}}$ by the
reverberation mapping of AGNs. The factors $f$ can be measured on the basis of two physical quantities, i.e. the gravitational redshifts
$z_{\rm{g}}$ and full widths at half maxima $v_{\rm{FWHM}}$ of broad lines. In the past it \textbf{has been} difficult to determine the
factors $f$ for individual AGNs. \textbf{We apply this new method to several} reverberation mapped Seyfert 1 galaxies. There is a correlation
between $f$ and broad-line region (BLR) radius $r_{\rm{BLR}}$, $f=5.4 r_{\rm{BLR}}^{0.3}$, for the gravitationally redshifted broad lines
He II, He I, H$\beta$ and H$\alpha$ in narrow-line Seyfert 1 galaxy (NLS1) Mrk 110. This correlation results from the radiation pressure
influence of \textbf{the} accretion disc on the BLR clouds. The radiation pressure influence seems to be more important than usually thought
in AGNs. Mrk 110 has $f \approx$ 8--16, distinctly larger than the mean $\langle f\rangle \approx 1$, usually used to estimate $M_{\rm{RM}}$
in the case of $v_{\rm{FWHM}}$. NGC 4593 and NLS1 Mrk 486 has $f\approx 3$ and $f\approx 9$, respectively. Higher $f$ \textbf{values} of several
tens are derived for \textbf{three other} NLS1s. There is a correlation between $f$ and accretion rate $\mathscr{\dot M}_{f=1}$,
$f=6.8\mathscr{\dot M}^{0.4}_{f=1}$ for five objects, where $\mathscr{\dot M}_{f=1}=\dot M_{\bullet}/L_{\rm{Edd}}c^{-2}$ \textbf{as $f=1$ is
assumed to estimate $M_{\rm{RM}}$ used in the Eddington luminosity $L_{\rm{Edd}}$}, $\dot M_{\bullet}$ is the mass accretion rate, and $c$
is the speed of light. These larger $f$ values will produce \textbf{higher} $M_{\rm{RM}}$ \textbf{values} and \textbf{lower} Eddington ratios.

\end{abstract}

\begin{keywords}
black hole physics -- galaxies: active -- galaxies: nuclei -- galaxies: Seyfert -- quasars: emission lines.
\end{keywords}

\section{INTRODUCTION}
Active galactic nuclei (AGNs), such as quasars and Seyfert galaxies, can be powered by the \textbf{release} of gravitational
potential energy of matter accreted onto supermassive black holes surrounded by accretion discs \citep{Re82,Re84}. The 
reverberation mapping (RM) model shows that the broad emission line variations are driven by the ionizing continuum variations 
through the photoionization process \citep[e.g.][]{BM82,Pe93}. Broad-line region (BLR) radius $r_{\rm{BLR}}$ can be determined 
by \textbf{the} time lag $\tau$ between the broad-line and continuum variations, as $r_{\rm{BLR}}=\tau c$, where $c$ is the 
speed of light. The RM observations and researches have been carried out for AGNs over the last several decades 
\citep[e.g.][]{Ka99,Ka00,Ka07,Pe05,De10,Ha11,Nu12,Du14,Du15,Pe14,Wa14,Ba15,Hu15}. Recently, some RM surveys \textbf{have been} 
proposed and carried out, such as the Sloan Digital Sky Survey (SDSS) spectroscopic RM project \citep{Sh15a,Sh15b,Sh16} and the
OzDES AGN spectroscopic RM project \citep{Ki15}. \textbf{These RM studies will be the most efficient method to accurately estimate 
black hole masses $M_{\rm{RM}}$ of AGNs at moderate-to-high redshifts. $M_{\rm{RM}}$ is given by $M_{\rm{RM}}=f v^2_{\rm{FWHM}} r_{\rm{BLR}}/G$,
where $f$ is the virial factors, $v_{\rm{FWHM}}$ is the full widths at half maxima of emission lines and $G$ is the gravitational 
constant \citep{Pe04}.} However, the virial factors $f$ in the $M_{\rm{RM}}$ estimates are uncertain due to the unclear kinematics 
and geometry of BLRs in AGNs \citep{Pe04,Wo15}. An average $\langle f\rangle \approx 1$ is derived by the black hole mass--stellar 
velocity dispersion ($M_{\bullet}-\sigma_{\ast}$) relation for the low redshift quiescent galaxies and/or reverberation mapped AGNs 
using $v_{\rm{FWHM}}$ of Balmer emission lines \citep{On04,Pi15,Wo15}. Constraining the virial factors $f$ is an important task for 
investigating black hole mass related issues.

The BLR cloud motions of the reverberation mapped AGNs are believed or assumed to be dominated by the gravitational forces
of the central supermassive black holes (i.e. virialized motions) \citep[e.g.][]{Kr91,Wa99,Kr01,Ba11,Du14,Wa14}. The virialized
motions generate the observed $v_{\rm{FWHM}}$ of optical broad emission lines, typically thousands of $\rm{km \/\ s^{-1}}$.
The optical BLRs usually span over distances of hundreds to thousands of the gravitational radii from the \textbf{central} black holes
for the reverberation mapped AGNs. \textbf{Despite} the huge distances, the broad lines should be redshifted by the central black hole's
gravity. Gravitational redshift in the weak field regime establishes pure shifts of spectral features without changing their
intrinsic shapes and in the strong field regime produces remarkable distortions of spectral shapes \citep{Mu06}. Remarkable
profile distortion is a key feature of relativistic spectra in AGNs with very skewed and asymmetric line profiles, e.g. iron
K$\alpha$ lines, generated in the emitting regions very close to the central black holes \citep[e.g.][]{Fa89,Po95,Ta95,Fa00,Re03}.
\textbf{In the literature}, there are three ways to measure the gravitational redshift $z_{\rm{g}}$. First, it is measured for a broad line
as a redshift difference with respect to \textbf{narrow emission lines like} [O {\sc iii}] $\lambda 5007$ \citep[e.g.][]{Zh90,Mc99b,Tr14}. 
Second, it is measured at different levels of the line intensity as a centroid shift with respect to the broad line peak, eliminating all 
spectra with blueshifted profiles \citep{Jo16}. Third, it is measured as a broad-line center shift of the root mean square (rms) spectrum 
with respect to the narrow line \citep{Ko03b}. A sign of $z_{\rm{g}}$ was found in a statistical sense for broad H$\beta$ in the
single-epoch spectra of over 20,000 quasars in the SDSS Data Release 7 (DR7) \citep{Tr14}. \citet{Jo16} found \textbf{a positive 
correlation between intrinsic redshift $\Delta z_{\rm{50}}$, dominantly caused by the gravitational effect, and $v_{\rm{FWHM}}$ of H$\beta$ 
for 209 AGNs taken from the SDSS DR7, which matches the theoretically expected relationship of $\Delta z_{\rm{50}}$ $\propto$ $v_{\rm{FWHM}}^2$.}
The redshifts of the rms profiles of broad lines and the BLR radii in Mrk 110 \textbf{also} follow the gravitational redshift prediction 
\citep[see Fig. 3 in][]{Ko03b}.

The RM masses of the black holes are $M_{\rm{RM}}\sim$ $10^{6}$--$10^{7}$ $M_{\odot}$ for Seyfert 1 galaxies \citep[e.g.][]{Ka99,
Ka00,Pe05,Be06,De10,Ha11,Du14,Du15,Wa14}. Seyfert 1 galaxies have a relatively high Eddington ratio $L_{\rm{bol}}/L_{\rm{Edd}}$,
where $L_{\rm{bol}}$ is the bolometric luminosity and $L_{\rm{Edd}}$ is the Eddington luminosity. Narrow-line Seyfert 1 galaxies
(NLS1s) seem to have a \textbf{higher} Eddington ratio. Some NLS1s \textbf{appear to be accreting at super-Eddington rates}. A large RM 
campaign was performed by the Yunnan Observatory 2.4 m telescope from 2012 to 2013 for AGNs with super-Eddington accreting massive black 
holes (SEAMBHs) \citep{Du14,Wa14}. \citet{Hu15} studied the properties of emission lines for 10 NLS1s with SEAMBHs, and found the redshifts
and/or blueshifts of H$\beta$ and Fe {\sc ii}. The inflow and outflow of BLR gas may generate redshifts and blueshifts, respectively.
\citet{Ko03a} ruled out that radial inflow or outflow motions are dominant in the BLR of NLS1 Mrk 110. \citet{Ko03b} found the 
gravitationally redshifted broad emission lines He II, He I, H$\beta$ and H$\alpha$. The BLRs will \textbf{"breathe"} with the central
radiation variations \citep[e.g.][references therein]{Ba15}. The breaths occur on short timescales of days to weeks in response to
continuum variations, and the broad-line shifts of $\sim$ 100 $\rm{km \/\ s^{-1}}$ were found over about one month. The breathing
effects of BLRs on the broad-line shifts might be eliminated in the \textbf{rms and mean spectra of the reverberation mapped AGNs by 
virtue of averaging over many breaths, and this was the approach employed in Mrk 110 by \citet{Ko03b}.} The redshifted H$\alpha$
and H$\beta$ broad lines were found with the rms spectra for Seyfert 1 galaxy NGC 4593, and their redward shifts might be interpreted
as the gravitational redshift \citep{Ko97}. 

\textbf{In this paper, we derive a new method to estimate the virial factors $f$ with the gravitationally redshifted broad optical emission lines 
in AGNs, and apply this new method to several reverberation mapped Seyfert 1 galaxies.} The structure of this paper is as follows. Section 2 
presents the method. Section 3 describes the application to Mrk 110. Section 4 is for the application to NGC 4593. Section 5 is for the 
applications to Mrk 493, \emph{IRAS} 04416 and Mrk 42. Section 6 presents discussion and conclusions. Throughout this paper, we use the 
standard cosmology with $H_0=70 \rm{\/\ km \/\ s^{-1} \/\ Mpc^{-1}}$, $\Omega_{\rm{M}}$ = 0.3, and $\Omega_{\rm{\Lambda}}$= 0.7 \citep{Sp03,Ri04}.

\section{METHOD}
The BLRs are distant from the central black holes for the reverberation mapped AGNs. The Schwarzschild metric will be
reasonable to describe the space-time around the BLRs. The Kerr metric and the Schwarzschild metric have the identical
effect on the gravitational redshift at distances larger than about one hundred gravitational radii from the black
holes \citep[see Fig. 9 in][]{Mu06}. The Schwarzschild space-time is
\begin{equation}
\begin{split}
\mathrm{d} s ^2 &=-g_{\rm{\mu \nu}}\mathrm{d} x^{\rm{\mu}} \mathrm{d} x^{\rm{\nu}} \\
&= \left(1- \frac{2GM_{\bullet}}{c^2 r} \right)c^2 \mathrm{d} t^2
- \left(1- \frac{2GM_{\bullet}}{c^2 r} \right)^{-1} \mathrm{d} r^2 \\
&\quad - r^2 \mathrm{d} \theta^2 -r^2 \sin^2 \theta \mathrm{d}
\varphi ^2,
\end{split}
\end{equation}
where $G$ is the gravitational constant and $M_{\bullet}$ is black hole mass. The ratio of the frequency of
atomic transition $\nu_{\rm{e}}$ at the BLR radius $r_{\rm{BLR}}$ to the frequency $\nu_{\rm{o}}$ observed at
infinite distance is (static cloud)
\begin{equation}
\begin{split}
\frac{\nu_{\rm{o}}}{\nu_{\rm{e}}}& = \frac{\left(-g_{00}
\right)^{1/2}_{r_{\rm{BLR}}}} {\left(-g_{\rm{00}}
\right)^{1/2}_{\infty}c} \\
&= \left( 1-\frac{2G M_{\bullet}}{c^2 r_{\rm{BLR}}} \right)^{1/2},
\end{split}
\end{equation}
where $(-g_{\rm{00}})_{\infty}= 1$ at the observer's frame. So, the gravitational redshift is
\begin{equation}
\begin{split}
z_{\rm{g}}&=\frac{\nu_{\rm{e}}}{\nu_{\rm{o}}} -1 \\
&=\left(1-\frac{2G M_{\bullet}}{c^2 r_{\rm{BLR}}}\right)^{-1/2}-1.
\end{split}
\end{equation}
The black hole mass $M_{\bullet}$ is
\begin{equation}
M_{\bullet}=\frac{1}{2}G^{-1}c^2 r_{\rm{BLR}} \left[ 1-\left(
1+z_{\rm{g}} \right) ^{-2} \right],
\end{equation}
and the first order approximation is
\begin{equation}
M_{\bullet}\cong G^{-1} c^2 z_{\rm{g}} r_{\rm{BLR}},
\end{equation}
if $z_{\rm{g}} \ll 1$. Equation (5) was used to estimate $M_{\bullet}$ in \citet{Ko03b} and \citet{Zh90}.

In general, the gravitational redshift $z_{\rm{g}}$ is derived from the redshift difference of the broad emission lines
relative to the narrow emission lines. \textbf{When} the narrow lines do not appear in the spectrum containing the broad lines,
the redshift difference $\Delta z_{\rm{i,j}}=z_{\rm{i}}-z_{\rm{j}}$ $=z_{\rm{g,i}}-z_{\rm{g,j}}$ for the broad lines
$i$ and $j$ \textbf{can be} used to estimate $M_{\bullet}$. Hence we have
\begin{equation}
\begin{split}
\Delta z_{\rm{i,j}} &=\left( 1-\frac{2G M_{\bullet}}{c^2
r_{\rm{BLR,i}}}\right)^{-1/2}- \left( 1-\frac{2G M_{\bullet}}{c^2
r_{\rm{BLR,j}}}\right)^{-1/2}\\
& \cong \frac{G M_{\bullet}}{c^2} \left( \frac{1}{r_{\rm{BLR,i}}}
- \frac{1}{r_{\rm{BLR,j}}}\right),
\end{split}
\end{equation}
where $r_{\rm{BLR}}\gg r_{\rm{g}}=GM_{\bullet}/c^2$ ($r_{\rm{g}}$ is the gravitational radius). And then we have
\begin{equation}
M_{\bullet}\cong G^{-1} c^2 \Delta z_{\rm{i,j}}  \left(
\frac{1}{r_{\rm{BLR,i}}} - \frac{1}{r_{\rm{BLR,j}}}\right)^{-1},
\end{equation}
where $r_{\rm{BLR,i}}$ and $r_{\rm{BLR,j}}$ correspond to the broad lines $i$ and $j$, respectively.
$M_{\bullet}$ can be measured by the virial theorem for the reverberation mapped AGNs \citep{Pe04}:
\begin{equation}
M_{\rm{RM}}=f\frac{v^2_{\rm{FWHM}} r_{\rm{BLR}}}{G}.
\end{equation}
If the RM method and the gravitational redshift method give the same mass $M_{\bullet}$, we have the virial factor
\begin{equation}
f=\frac{1}{2} \frac{c^2}{v^2_{\rm{FWHM}} } \left[ 1-\left(1+z_{\rm{g}} \right) ^{-2} \right],
\end{equation}
for the reverberation mapped AGNs. This gravitational redshift approach will be a \textbf{simple and direct} method to estimate the virial
factor $f$ in equation (8). If $f$ is fully generated by the oblique effect of \textbf{a disc-like BLR} with inclinations
$\theta$, $f= 1/4\sin^2 \theta$ \citep{Mc01}. \textbf{For} $f=1$, $\theta =30$ degrees. Hereafter, $M_{\rm{grav}}$ denotes
$M_{\bullet}$ estimated by equations (4), (5) and (7), and $M_{\rm{RM}}$ denotes $M_{\bullet}$ estimated by equation (8).

\section{APPLICATION TO MRK 110}
The shifts of the rms line centers (uppermost 20\%) with respect to the narrow lines are identified as the gravitational
redshifts \citep{Ko03b}. The mean profiles of the broad lines are not shifted with respect to the forbidden narrow lines
on the other hand. Therefore, the differential shifts of the rms profiles with respect to the narrow lines are identical to
their shifts with respect to the mean profiles \citep{Ko03b}. The virial factors $f$ are derived by equation (9) for the
broad lines He II, He I, H$\beta$ and H$\alpha$ (see Table 1). Considering the errors of $v_{\rm{FWHM}}$ and $z_{\rm{g}}$,
the $f$ distribution is generated with equation (9) by $10^4$ realizations of Monte Carlo simulation. The mean and standard
deviation of this distribution are regarded as the expectation and uncertainty of $f$, respectively. The helium and hydrogen
lines have $f\approx$ 8--16 that are larger than the average $\langle f\rangle \approx 1$ usually accepted for the reverberation
mapped AGNs. These $f$ values of the helium lines are slightly smaller than those of the hydrogen lines. If $f$ is fully generated
by the oblique effect of the disc-like BLRs, $\theta \approx$ 7--10 degrees (see Table 1), which confirm the nearly face-on view
of accretion disc in Mrk 110 suggested by \citet{Ko03a,Ko03b}. The broad lines He II, He I, H$\beta$ and H$\alpha$ have obvious
stratification in the BLRs, as predicted by the virial theorem \citep{Ko03b}, and \textbf{arise from increasing distances} from the
central black hole. These stratification BLRs are dominated by the central supermassive black hole's gravity. Mrk 110 will have
the virial velocity $v^2_{\rm{c}}=G M_{\bullet} r^{-1}_{\rm{BLR}}$ and $v^2_{\rm{FWHM}}\propto v^2_{\rm{c}} \propto r^{-1}_{\rm{BLR}}$.
We have $f \cong z_{\rm{g}} c^2/v^2_{\rm{FWHM}}$ because $z_{\rm{g}} \ll 1$ in equation (9) for Mrk 110, and then $f \propto z_{\rm{g}}
r_{\rm{BLR}}\propto M_{\bullet}$ (see equation [5]). $M_{\bullet}$ can be regarded as a constant in observation periods for individual
AGNs. So, $f$ \textbf{should be} independent of $r_{\rm{BLR}}$. However, there is a positive correlation between $f$ and $r_{\rm{BLR}}$ 
for Mrk 110,
\begin{equation}
\log f=0.73(\pm 0.08)+0.30(\pm 0.07)\times \log r_{\rm{BLR}}
\end{equation}
\textbf{with a Pearson's correlation coefficient $r=0.948$ at the confidence level of 94.8 percent. The ''FITEXY'' estimator \citep{Pr92}
gives $\log f=0.74(\pm 0.21)+0.29(\pm 0.22)\times \log r_{\rm{BLR}}$ for the data and uncertainties in x and y with a chi-square 
$\chi ^2=0.180$ and a goodness-of-fit $Q= 0.914$ (see Figure 1). These two best fit lines are indistinguishable.} This correlation of 
$f=5.4(\pm1.0)\times r_{\rm{BLR}}^{0.3\pm 0.1}$ is inconsistent with the independent prediction of the virial theorem.

There are two paths to estimate $M_{\bullet}$, i.e. the single broad-line estimation (equation [4]) and the broad-line-to-broad-line
comparison (equation [7]). The gravitational masses $M_{\rm{grav}}$ are estimated for the broad lines He II, He I, H$\beta$ and
H$\alpha$ (see Tables 1 and 2). According to basic error propagation conventionally used in the RM method, we estimate the uncertainty
of $M_{\rm{grav}}$ from the errors of $r_{\rm{BLR}}$ and $z_{\rm{g}}$. Each $M_{\rm{grav}}$ in Table 2 will correspond to two
$M_{\rm{grav}}$ in Table 1, because equation (7) is based on two broad lines. Comparisons show that the masses estimated by these two
paths are consistent with each other (see Figure 2). This agreement indicates that the broad-line-to-broad-line comparison path is
feasible and reliable to estimate $M_{\bullet}$. The black hole masses in Tables 1 and 2 are also consistent with the mean mass of
$\log M_{\rm{grav}}/ M_{\rm{\odot}}=8.15$ derived in \citet{Ko03b}. The broad-line-to-broad-line comparison path \textbf{avoids any 
potential difficulty coming from blueshifts of narrow [O {\sc iii}] $\lambda 5007$ in the single broad-line path. Such narrow line blueshifts 
would} equivalently generate the broad line redshift when the blueshifted narrow line is used as the reference to estimate $z_{\rm{g}}$.
In principle, the broad-line-to-broad-line comparison path could mostly eliminate the line shift influence due to the BLR "breath".
The broad-line-to-broad-line comparison path seems to be better than the single broad-line path to estimate $M_{\bullet}$ \textbf{when 
there are several broad lines in the spectrum.} So, equation (7) is appropriate to estimate $M_{\bullet}$ for multi-broad-line AGNs. Equation (4) is
appropriate for a single broad line except for the blueshift issue of the narrow line used as the reference to estimate $z_{\rm{g}}$.
\begin{table*}
\centering
\begin{minipage}{155mm}
\caption{The details of $v_{\rm{FWHM}}$, $z_{\rm{g}}$, $\tau$, $M_{\rm{grav}}$, $f$ and $\theta$ for Mrk 110 and NGC 4593
\label{tbl-1}}

\begin{tabular}{cccccccc}
\hline \hline

Object & Line & $\frac{v_{\rm{FWHM}}}{\rm{km \/\ s^{-1}}}$ & $z_{\rm{g}}$
& $\tau$ (days)& $\log \frac{M_{\rm{grav}}}{M_{\odot}}$ & $f$ & $\theta^{\circ}$  \\
(1)&(2)&(3)&(4)&(5)&(6)&(7)&(8) \\

\hline

Mrk 110 & He II $\lambda$4686 &$4444 \pm 200$ &0.00180$\pm$0.00020 &$3.9\pm 2$&8.09$\pm$0.23 &8.23$\pm$1.18 &10.12$\pm$0.74 \\

Mrk 110 & He I $\lambda$5876 &$2404 \pm 100$ &0.00062$\pm$0.00020 &$10.7 \pm6 $&8.07$\pm$0.28 &9.70$\pm$3.21 &9.24$\pm$1.56 \\

Mrk 110 & H$\beta$ $\lambda$4861&$1515 \pm 100$ &0.00039$\pm$0.00017&$24.2 \pm 4$&8.22$\pm$0.20 &15.65$\pm$6.79 &7.26$\pm$1.68 \\

Mrk 110 & H$\alpha$ $\lambda$6563 &$1315 \pm 100$ &0.00025$\pm$0.00017&$32.3 \pm 5$ &8.15$\pm$0.30 &14.59$\pm$8.27 &7.52$\pm$2.79 \\

\hline

NGC 4593 & H$\alpha$ $\lambda$6563 &$3400 \pm 200$ &0.00037$\pm$0.00010&$3.8 \pm 2.0$ &7.39$\pm$0.26 &2.91$\pm$0.86 &17.04$\pm$2.60 \\
\hline

\end{tabular}
\\Notes: Column 1: object names; Column 2: emission line names; Column 3: $v_{\rm{FWHM}}$ of broad lines in the rms spectra;
Column 4: the gravitational redshifts; Column 5: time lags of lines in the rest frame; Column 6: the gravitational masses derived
from equation (4); Column 7: the virial factors; Column 8: inclinations if $f$ is fully generated by the inclination effect of
the disc-like BLRs.

\end{minipage}
\end{table*}
\begin{table}
\centering
\begin{minipage}{70mm}
\caption{$M_{\rm{grav}}$ given by broad line pairs for Mrk 110
\label{tbl-2}}

\begin{tabular}{cccc}
\hline \hline

Line & He I $\lambda$5876 & H$\beta$ $\lambda$4861 & H$\alpha$ $\lambda$6563  \\
(1)&(2)&(3)&(4) \\

\hline

He II $\lambda$4686  &8.10$\pm$0.39 & 8.06$\pm$0.28 & 8.08$\pm$0.26  \\

He I $\lambda$5876 &                       & 7.89$\pm$0.66 & 8.02$\pm$0.48  \\

H$\beta$ $\lambda$4861 &                    &                       & 8.38$\pm$0.82  \\

\hline

\end{tabular}
\\Notes: the gravitational masses are derived from equation (7) and scaled as  $\log \frac{M_{\rm{grav}}}{M_{\odot}}$.

\end{minipage}
\end{table}
\begin{figure}
\begin{center}
\includegraphics[angle=-90,scale=0.28]{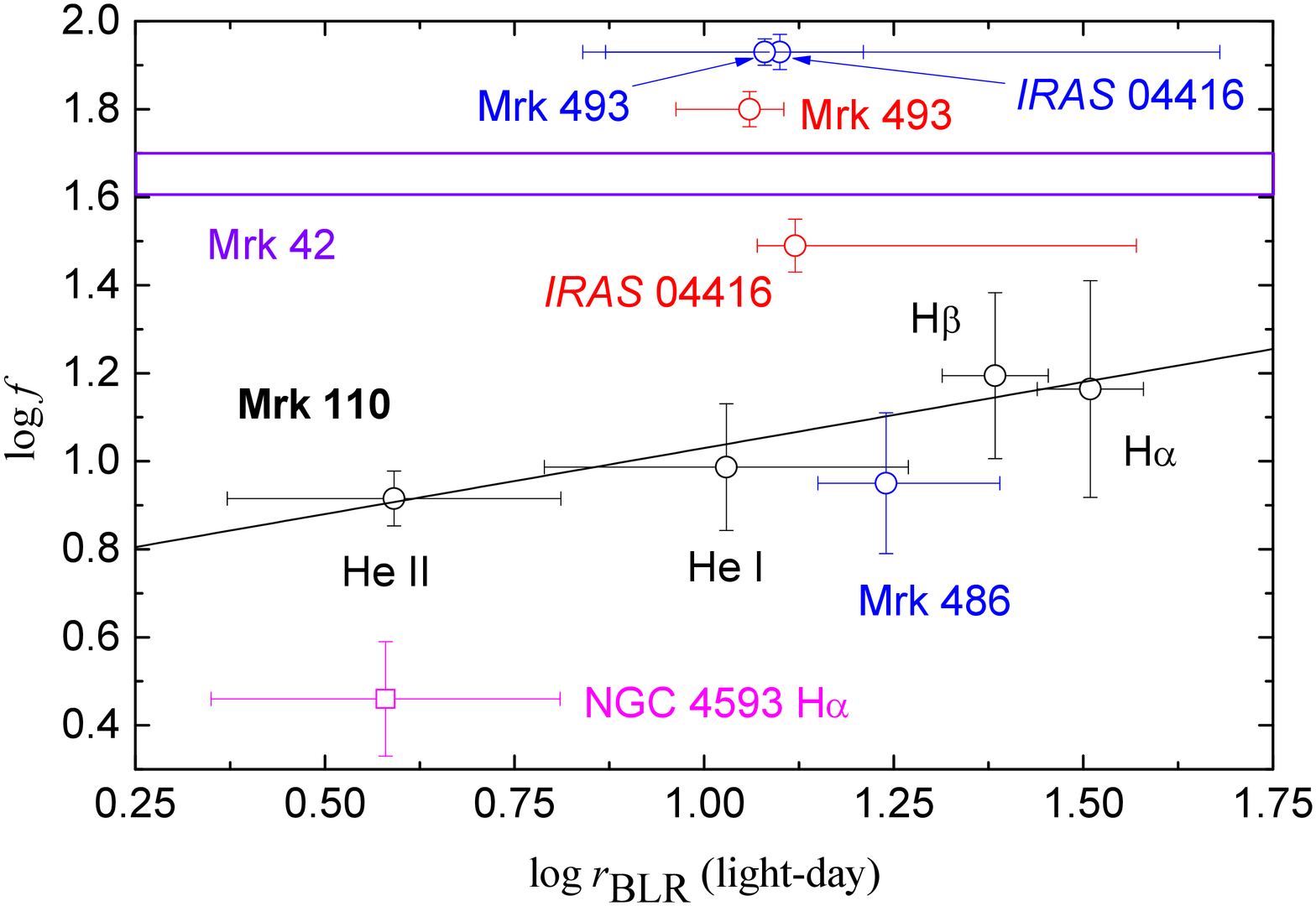}
\end{center}
 \caption{$\log f$ vs $\log r_{\rm{BLR}}$. Black circles are for Mrk 110. Circles in color are for SEAMBH AGNs. Red circles
 denote the H$\beta$ line, and blue circles denote the Fe {\sc ii} line. Violet lines denote the $f$ values for Mrk 42. Black line
 is the best fit line to the black circles.}
  \label{fig1}
\end{figure}
\begin{figure}
\begin{center}
\includegraphics[angle=-90,scale=0.28]{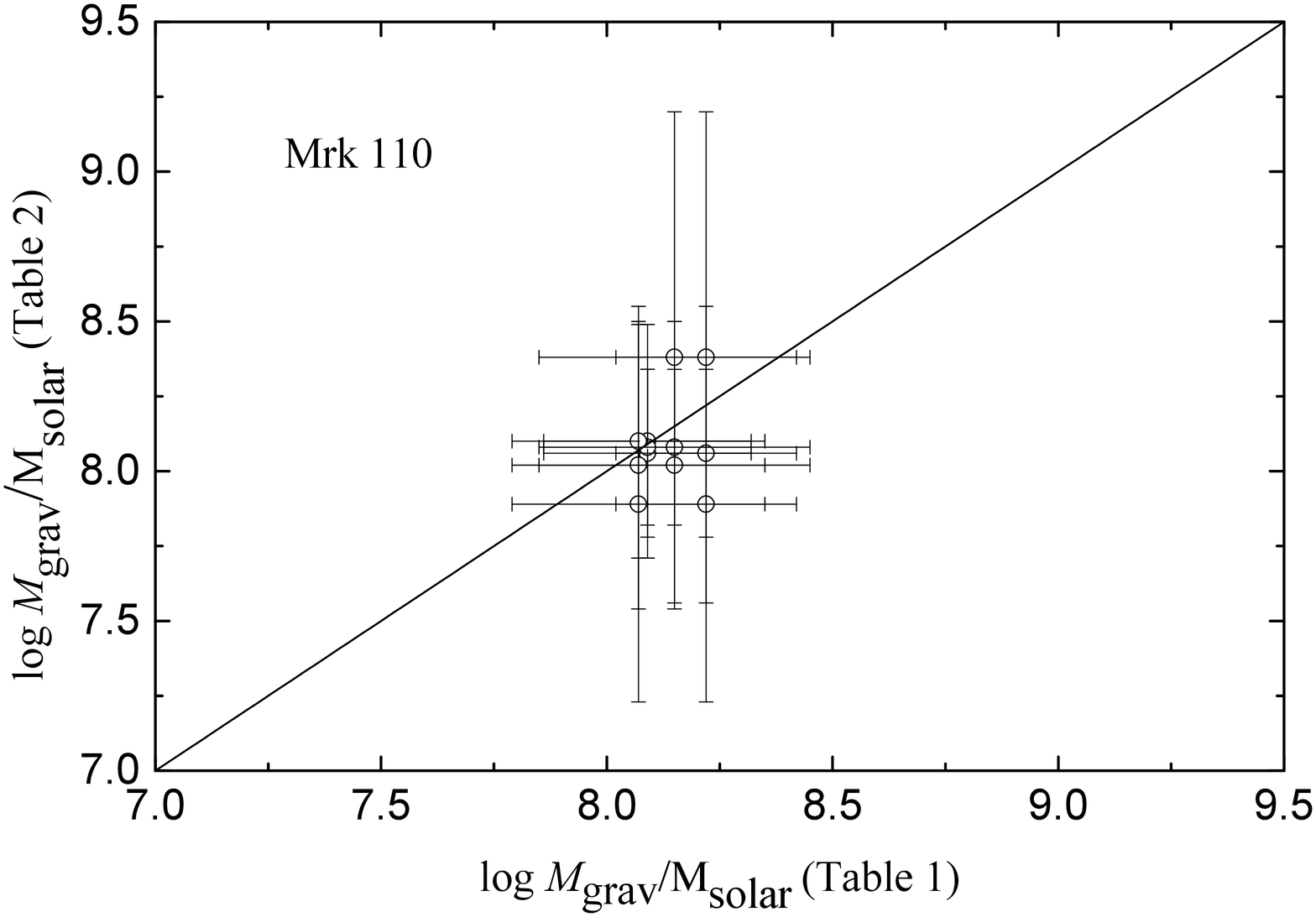}
\end{center}
 \caption{Comparison between $M_{\rm{grav}}$ derived from equations (4) and (7). Black line is $y=x$.}
  \label{fig2}
\end{figure}

\section{APPLICATION TO NGC 4593}
NGC 4593 is a nearby Seyfert 1 galaxy. This object is a sub-Eddington accreting AGN with an average $\mathscr{\dot M}=0.08$,
where $\mathscr{\dot M}=\dot M_{\bullet}/L_{\rm{Edd}}c^{-2}$ is the dimensionless accretion rate and $\dot M_{\bullet}$
is the mass accretion rate \citep{Du15}. \textbf{Hereafter, $\mathscr{\dot M}$ is denoted by $\mathscr{\dot M}_{f=1}$, because
its calculation is assuming $f=1$ to estimate $M_{\rm{RM}}$ used in $L_{\rm{Edd}}$.} Early RMs yielded a lag of four days
indicating a very compact BLR \citep{Di94}. Recent RMs confirm this lag of four days \citep[e.g.][references therein]{Du15}.
\citet{Ko97} found the redshifted H$\alpha$ and H$\beta$ broad lines with the rms spectra, and thought that the observed redshift
might be interpreted as the gravitational redshift. The central component of H$\alpha$ is redshifted by $110\pm 30$ $\rm{km}\/\ s^{-1}$
with respect to the narrow line, and the H$\beta$ rms spectrum shows a redshift of the same order of magnitude. We have $v_{\rm{FWHM}}$
$=3400\pm 200$ $\rm{km}\/\ s^{-1}$ for both of the H$\alpha$ rms and mean spectra. The relevant physical quantities of H$\alpha$ are
listed in Table 1. $M_{\rm{grav}}=2.5(\pm 1.5)\times 10^{7}$ $M_{\odot}$ is given by equation (4). $f=2.9$ is given by the redshifted
H$\alpha$ line. This $f$ value is slightly higher than $\langle f\rangle \approx 1$, and is smaller than those of Mrk 110.

In plot of $\log f$ versus $\log r_{\rm{BLR}}$, NGC 4593 is below the best fit line to Mrk 110 (see Figure 1). If $f$ is fully 
generated by the inclination effect of the BLR for NGC 4593, $\theta \approx$ 17 degrees. The inclination effect will generate 
different $f$ even for the same BLR \citep[e.g.][]{Mc01}. Another possible explanation of the different $f$ is that the accretion 
rate $\mathscr{\dot M}_{f=1}$ dominates the factor $f$, because $f=2.9$ and $\mathscr{\dot M}_{f=1}=0.08$ in NGC 4593 are smaller 
than $f=14.6$ and $\mathscr{\dot M}_{f=1}=5.89$ \textbf{\citep{Du15}} in Mrk 110 for H$\alpha$, respectively. A new quantity 
$\mathscr{\dot M}_{\rm{f}}$ is defined as a normalization of $\mathscr{\dot M}$ relative to $f$. $\mathscr{\dot M}_{\rm{f}}$ 
corrects the underestimation of $M_{\rm{RM}}$ due to the \textbf{use} of $f=1$, takes into account differences in the $\mathscr{\dot M}$ 
estimates, and represents the more \textbf{realistic} dimensionless accretion rate. NGC 4593 and Mrk 110 have $\mathscr{\dot M}_{\rm{f}}$ 
$=$ 0.03 and 0.40 for H$\alpha$, respectively. The normalized value $\mathscr{\dot M}_{\rm{f}}$ reflects the composite effect of 
the central black hole's gravity, the central radiation pressure and the kinematics and geometry of BLR. Mrk 110 has higher 
$\mathscr{\dot M}_{\rm{f}}$ and $f$ than NGC 4593.

\section{APPLICATIONS TO Mrk 493, \emph{IRAS} 04416 and Mrk 42}
\citet{Hu15} studied the properties of emission lines for 10 NLS1s. Mrk 493, \emph{IRAS} 04416+1215 and Mrk 42 have the
redward shifted optical Fe {\sc ii} and H$\beta$ lines with respect to [O {\sc iii}] $\lambda 5007$. The virialized motions
of BLR clouds are assumed to estimate $M_{\rm{RM}}$ in the RM of H$\beta$ for these NLS1s \citep{Du14,Wa14}. The redward
shifts of H$\beta$ and Fe {\sc ii} may be the gravitational redshifts. The Fe {\sc ii} lag is consistent with the H$\beta$ lag
for the same source (see Table 3). Mrk 493 and Mrk 42 have the Fe {\sc ii} $v_{\rm{FWHM}}$ indistinguishable from the H$\beta$
$v_{\rm{FWHM}}$. The Fe {\sc ii} $v_{\rm{FWHM}}$ is slightly smaller than the H$\beta$ $v_{\rm{FWHM}}$ in \emph{IRAS} 04416.
There are no obvious stratification \textbf{for these two emission lines} in the BLRs for Mrk 493, \emph{IRAS} 04416 and Mrk 42. 
Here, we regard these redward shifts of the Fe {\sc ii} and H$\beta$ lines as the gravitational redshifts. The details for these 
three NLS1s are listed in Table 3. The factors $f$ and the masses $M_{\rm{grav}}$ are estimated by equations (9) and (4), respectively. 
These three NLS1s have higher $f$ than Mrk 110 (see Figure 1). Their $f$ values are much larger than $\langle f\rangle \approx 1$,
\textbf{implying larger black hole masses than typically estimated from $M_{\rm{RM}}$}. The higher
masses will decrease the Eddington ratios. These results are very similar to those of Mrk 110 with a mean $f\approx 12$, which
makes $M_{\rm{RM}}$ increase by an order of magnitude.

These three NLS1s and Mrk 110 have comparable $r_{\rm{BLR}}$ and $M_{\rm{grav}}$, but distinctly different $f$ (see Figure 1).
These behaviors may be from the difference of accretion states. Mrk 493 and \emph{IRAS} 04416 have $\mathscr{\dot M}_{f=1}=75.9$ and
$\mathscr{\dot M}_{f=1}=426.6$, respectively \citep{Du15}. Mrk 493 and \emph{IRAS} 04416 have a mean $f=73.3$ and $f=58.7$, respectively.
Thus, Mrk 493 and \emph{IRAS} 04416 have $\mathscr{\dot M}_{\rm{f}}=1.04$ and $\mathscr{\dot M}_{\rm{f}}=7.27$, respectively. Mrk 110
has a mean $f=12.0$ and $\mathscr{\dot M}_{\rm{f}}=0.49$. In terms of $\mathscr{\dot M}_{f=1}$ and $\mathscr{\dot M}_{\rm{f}}$, Mrk 493
and \emph{IRAS} 04416 are at higher accretion states than Mrk 110. The higher accretion rate seems to result in the larger $f$.
\textbf{The higher accretion rate will result in stronger radiation pressure on the BLR clouds.} The higher radiation pressure will counteract
a \textbf{greater} proportion of the black hole's gravity, and then the BLR clouds will have a smaller virial velocity, i.e. $v_{\rm{FWHM}}$ 
will decrease if the radiation pressure increases, given that the BLR clouds are in virialized motions. Ultimately, the radiation pressure
will influence $M_{\rm{RM}}$. Thus, the factors $f$ given by equation (9) will be higher than expected from the virial assumption
without considering the radiation pressure.

A higher mass $\log M_{\rm{grav}}/M_{\rm{\odot}}\approx 9.5$ is estimated by equation (7) for \emph{IRAS} 04416. A very
large error of $\log M_{\rm{grav}}/M_{\rm{\odot}}$ will be generated by the larger error of $\tau$ ($r_{\rm{BLR}}=c\tau$)
(see Table 3). This larger mass is on the same order of magnitude as the mass estimated by equation (4) for Fe {\sc ii} when
considering its errors. This means that the single broad-line path and the broad-line-to-broad-line comparison path are equally
applicable to estimating $M_{\bullet}$ for \emph{IRAS} 04416. A futher test is needed in the future with high quality time
lags of H$\beta$ and Fe {\sc ii} to check the validity of these two paths. The current results indicate that the redward shifts
of H$\beta$ and Fe {\sc ii} with respect to [O {\sc iii}] are reliable to be used as the gravitational redshifts, i.e. the
[O {\sc iii}] redshfit may be regarded as the systemic redshift. Equation (7) cannot be applied to Mrk 493. The applications
to Mrk 110 and \emph{IRAS} 04416 show that the broad-line-to-broad-line comparison path is more applicable to those AGNs with
obvious stratification in the BLRs, such as Mrk 110.

\begin{table*}
\centering
\begin{minipage}{155mm}
\caption{The details of $v_{\rm{FWHM}}$, $z_{\rm{g}}$, $\tau$, $M_{\rm{grav}}$, $f$ and $\theta$ of NLS1s for the H$\beta$ and
Fe{\sc ii} lines
\label{tbl-3}}

\begin{tabular}{cccccccc}
\hline \hline

Object& Line & $\frac{v_{\rm{FWHM}}}{\rm{km \/\ s^{-1}}}$ & $z_{\rm{g}}$
& $\tau$ (days)& $\log \frac{M_{\rm{grav}}}{M_{\odot}}$ & $f$ & $\theta^{\circ}$  \\
(1)&(2)&(3)&(4)&(5)&(6)&(7)&(8) \\

\hline

\emph{IRAS} 04416+1215 & H$\beta$ & 1522$\pm$44 & 0.00080$\pm$0.00010 & $13.3^{+13.9}_{-1.4}$ & $8.27^{+0.46}_{-0.07}$ & 31.14$\pm$4.29 & 5.14$\pm$0.37 \\

Mrk 42 & H$\beta$ & 802$\pm$18 & 0.00029$\pm$0.00006 &      &    & 40.60$\pm$8.60 & 4.50$\pm$0.55 \\

Mrk 493 & H$\beta$  & 778$\pm12$ & 0.00042$\pm$0.00004 & $11.6^{+1.2}_{-2.6}$ & $7.93^{+0.06}_{-0.11}$ & 62.44$\pm$6.26 & 3.63$\pm$0.19 \\

Mrk 486 & H$\beta$  & 1942$\pm67$ & -0.00015$\pm$0.00003 & $23.7^{+7.5}_{-2.7}$ &  &   &   \\

\hline

\emph{IRAS} 04416+1215 & Fe {\sc ii} & 1313$\pm$50 & 0.00165$\pm$0.00007 & $12.6^{+16.7}_{-6.7}$ & $8.56^{+0.58}_{-0.23}$ & 86.31$\pm$7.56 & 3.09$\pm$0.14 \\

Mrk 42     & Fe {\sc ii}   & 787$\pm$16 & 0.00035$\pm$0.00006 &    &    & 50.94$\pm$8.97 & 4.02$\pm$0.38 \\

Mrk 493  & Fe {\sc ii}   & 780$\pm$9 & 0.00057$\pm$0.00003 & $11.9^{+3.6}_{-6.5}$ & $8.08^{+0.13}_{-0.24}$ & 84.25$\pm$4.85 & 3.12$\pm$0.09 \\

Mrk 486 & Fe {\sc ii}   & 1790$\pm88$ & 0.00032$\pm$0.00011 & $17.3^{+5.8}_{-3.7}$ & $7.99^{+0.21}_{-0.18}$ & 8.98$\pm$3.21 & 9.60$\pm$1.73  \\

\hline

\end{tabular}
\\Notes: Same as in Table 1 except for Column 3: the H$\beta$ $v_{\rm{FWHM}}$ from the mean spectra, and the Fe {\sc ii} $v_{\rm{FWHM}}$ given by
 the mean and standard deviation obtained from the measurements of individual-night spectra \citep{Hu15}.

\end{minipage}
\end{table*}

\section{DISCUSSION AND CONCLUSIONS}
The gravitational redshift effect will exist in the broad lines as long as the BLRs surround the central supermassive black holes.
However, this gravitational effect is not always observable due to several factors. The first factor is the observational accuracies,
such as the low signal to noise ratios. The second is the breathing of BLRs that shifts the broad lines redward or blueward. The rms
and mean spectra may eliminate the breathing effect of BLRs on $z_{\rm{g}}$. The $z_{\rm{g}}$ and $r_{\rm{BLR}}$ of broad lines follow
the gravitational redshift prediction in Mrk 110 \citep[see Fig. 3 in][]{Ko03b}, indicating that the broad line shifts are dominated
by the gravitational redshifts for the rms spectra. The rms spectra were used to measure $z_{\rm{g}}$ of the broad lines in Mrk 110
\citep{Ko03b} and NGC 4593 \citep{Ko97}. The third is that the inflow may also produce the broad line redshifts relative to the narrow
lines. The RM method assumes that the BLR cloud motions meet the virial theorem. It was ruled out that radial inflow or outflow motions
are dominant in the BLRs of NGC 4593 \citep{Ko97}, Mrk 110 \citep{Ko03a}, Mrk 50 \citep{Ba11} and Mrk 1044 \citep{Du16}. The fourth is
the other broad and/or narrow lines blended with the target broad line. These blended components will influence the $v_{\rm{FWHM}}$ and
centroids of broad lines in the rms spectra \citep[e.g.][]{Ba15}, and may make it difficult to measure $z_{\rm{g}}$.

\citet{Jo16} eliminated all spectra with the blueshifted profiles. Some other effects (e.g. outflows) could be more dominant
than the gravitational effects in these spectra, which are not convenient for the researches on $z_{\rm{g}}$. It is difficult
to state that all spectra with strong outflow influence are completely eliminated, since the possible combination of the outflows
and the gravitational effects generates symmetrical line shape \citep{Jo16}. Their sample number is about 1\% of that in \citet{Tr14}.
So, it is difficult to measure $z_{\rm{g}}$ due to some other effects, and the chance of finding $z_{\rm{g}}$ is very low.
The applications may be influenced by the possible combination. The broad-line-to-broad-line comparison path may mostly eliminate
the underlying effects (inflows or outflows) by the using of different broad lines rather than a single broad line. These masses
derived from equations (4) and (7) for Mrk 110 are consistent with each other within the uncertainties (see Figure 2). For 
\emph{IRAS} 04416, these masses given by equation (7) are larger than those given by equation (4), but they are
on the same order of magnitude considering the uncertainties. This implies that the underlying effects are notable, but weaker than
the gravitational redshift effect for \emph{IRAS} 04416, and should be mostly eliminated by equation (7). Though, equations (4), (5)
and (9) are not applicable to the blueshifted lines (the blueshift is not eliminated for the single broad line), equation (7) may be
applicable. NLS1 Mrk 486 has the redshifted Fe {\sc ii} broad line (redshifted on $\pm 1 \sigma$) and the blueshifted H$\beta$ broad
line (blueshifted on $\pm 1 \sigma$) \citep[see Table 2 in][]{Hu15}. Mrk 486 has $\log M_{\rm{grav}}/M_{\odot}=$ $8.72^{+0.66}_{-0.38}$
given by equation (7) for Fe {\sc ii} and H$\beta$, and $\log M_{\rm{grav}}/M_{\odot}=$ $7.99^{+0.21}_{-0.18}$ given by equation (4)
for Fe {\sc ii}. These two masses are on the same order of magnitude. The outflow effect may be not strong, but more notable relative
to the gravitational redshift effect in Mrk 486. $f \approx$ 9 is estimated by Fe {\sc ii} for Mrk 486 (see Table 3). This value might
be the lower limit of $f$, for that the redshifted Fe {\sc ii} may not be purely due to the gravitational redshift effect.

In the absence of the radiation pressure on the BLR clouds of AGNs with $M_{\bullet}$, there is $v^2_{\rm{c}}=G
M_{\bullet}/r_{\rm{BLR}}=r^{-1}_{\rm{BLR}}(r_{\rm{g}}) c^2 $, where $r_{\rm{BLR}}(r_{\rm{g}})$ is in units of $r_{\rm{g}}$.
At the same time, $v^2_{\rm{FWHM}}\propto v^2_{\rm{c}} \propto r^{-1}_{\rm{BLR}}(r_{\rm{g}})$. Equations (4) and (9) are
combined to give $f=G M_{\bullet}/(r_{\rm{BLR}} v^2_{\rm{FWHM}})\propto r^{-1}_{\rm{BLR}}(r_{\rm{g}}) / v^2_{\rm{FWHM}}\propto C$,
where $C$ is independent of $r_{\rm{BLR}}$ and $M_{\bullet}$. However, \textbf{we found in Mrk 110 that $f=5.4 r_{\rm{BLR}}^{0.3}$ 
(see Figure 1). This discrepancy may result from having ignored the radiation pressure from the accretion disc on the BLR clouds.} 
The radiation pressure will push these clouds towards the larger radius, compared to that in the absence of the radiation pressure. 
The influence of radiation pressure on RM was studied in depth by \citet{Ma08}, in which the absorption of ionizing photons
and the scattering of nonionizing photons are combined to generate the radiation pressure for NLS1s. After the radiation pressure
correction, NLS1s have large black hole masses similar to other broad-line AGNs and follow the same $M_{\bullet}-\sigma_{\ast}$
relation as other active and normal galaxies \citep{Ma08}. This means that $f$ is larger than unity, due to considering the radiation 
pressure. The high-$f$ value is consistent with our results for NLS1s. \textbf{\citet{Ma08} take the radiation pressure force $F_{\rm{r}}$ 
as being proportional to $r^{-2}_{\rm{BLR}}$}, and the impact of radiation pressure then is to reduce the effective black hole mass 
by a fixed value for all emission lines in the BLR. In contrast, we parameterise the impact of radiation pressure with a radial power-law. 
\textbf{In the BLR model of \citet{Ne90}, the BLR cloud pressure is assumed to be determined by the pressure of the external medium, which 
has a simple power-law radial profile. \citet{Ka99} found a good fit to the emission line responses in NGC 5548 when the external pressure 
(and thus also the BLR gas density) was proportional to $r^{-1}_{\rm{BLR}}$. In such a model, the cloud cross section $\sigma_{\rm{c}}$ is 
then proportional to $r^{2/3}_{\rm{BLR}}$, and $F_{\rm{r}}$ is proportional to $r_{\rm{BLR}}^{-2} \sigma_{\rm{c}} \propto r_{\rm{BLR}}^{-4/3}$.} 
The effective black hole's gravity is $F^{\rm{eff}}_{\rm{g}}=F_{\rm{g}}-F_{\rm{r}}=A r_{\rm{BLR}}^{-2}-Br_{\rm{BLR}}^{-4/3}=E $
$r_{\rm{BLR}}^{-(2+\alpha)}$, \textbf{where $r_{\rm{BLR}}^{-\alpha}$ is the correction factor from the central disc radiation and $\alpha >0$}. \textbf{$Ar_{\rm{BLR}}^{-2}-Br_{\rm{BLR}}^{-4/3}=0$ gives a resolution $r_{\rm{gr}}$, and $E r_{\rm{BLR}}^{-(2+\alpha)}$ is positive for 
$r<r_{\rm{gr}}$. $F^{\rm{eff}}_{\rm{g}}$ can not be modelled as $E r_{\rm{BLR}}^{-(2+\alpha)}$ around $r_{\rm{gr}}$. Considering the centrifugal 
force $F_{\rm{c}}$ of a cloud, there is an equilibrium point $r_{\rm{grc}}$, and $r_{\rm{grc}}$ is smaller than $r_{\rm{gr}}$. For comparable 
$F_{\rm{g}}$, $F_{\rm{r}}$ and $F_{\rm{c}}$, $r_{\rm{gr}}$ should be several times of $r_{\rm{grc}}$. For BLR clouds, there are series of 
$r_{\rm{grc}}$. $F^{\rm{eff}}_{\rm{g}}$ can be effectively modelled as $E r_{\rm{BLR}}^{-(2+\alpha)}$ for these $r_{\rm{grc}}$.} Therefore, 
there is a ratio of $F_{\rm{g}}/F^{\rm{eff}}_{\rm{g}}\propto r^{\alpha}_{\rm{BLR}}$. We have $f=M_{\bullet}/M_{\rm{RM}}(f=1)=$
$M_{\rm{grav}}/M_{\rm{RM}}(f=1)\propto F_{\rm{g}}/ F^{\rm{eff}}_{\rm{g}}\propto r^{\alpha}_{\rm{BLR}}$, again. \textbf{At the other hand}, 
there is $v^2_{\rm{c}}\neq G M_{\bullet}/r_{\rm{BLR}}$ in the presence of the radiation pressure. We have $v^2_{\rm{cr}}=G M_{\bullet}
/r_{\rm{BLR}}^{1+\alpha}$ instead of $v^2_{\rm{c}}= G M_{\bullet}/r_{\rm{BLR}}$. So, $v^2_{\rm{cr}} < v^2_{\rm{c}}$ and $v^2_{\rm{FWHM}}\propto
v^2_{\rm{cr}} \propto r^{-1}_{\rm{BLR}}(r_{\rm{g}}) r_{\rm{BLR}}^{-\alpha}$. Thus, $f\propto r^{-1}_{\rm{BLR}}(r_{\rm{g}}) /
v^2_{\rm{FWHM}} \propto r_{\rm{BLR}}^{\alpha}$, i.e. $\log f =D + \alpha \log r_{\rm{BLR}}$, where $D$ is independent of $r_{\rm{BLR}}$ 
and $M_{\bullet}$. The virial factor $f$ is a function of $r_{\rm{BLR}}$ rather than $r_{\rm{BLR}}(r_{\rm{g}})$.
Thus, the interpretation of the effects of radiation pressure is reasonable. 

Parameter $\alpha$ is likely different from one to another AGN and might depend on accretion rate. We have $\alpha = 0.3$ for Mrk 110
(see Figure 1). The lines parallel to the best fit line to Mrk 110 cannot connect two points of Mrk 493 or \emph{IRAS} 04416 even
considering the corresponding errors. Mrk 110, Mrk 493 and \emph{IRAS} 04416 have $\mathscr{\dot M}_{f=1}=$ 5.89, 75.9 and 426.6 or 
$\mathscr{\dot M}_{\rm{f}}=$ 0.49, 1.04 and 7.27, respectively. \textbf{Mrk 493 and \emph{IRAS} 04416 have larger accretion rates than 
Mrk 110 and it is possible that $\alpha$ depends on the accretion rate, but the present data do not allow us to test this hypothesis.} 
There is a \textbf{strong} positive correlation between $f$ and $\mathscr{\dot M}_{f=1}$, $f=6.8(\pm 1.6)\mathscr{\dot M}_{f=1}^{0.4\pm 0.1}$
(see Figure 3), which \textbf{is consistent with $f$ being dominated} by the radiation pressure. This correlation \textbf{supports} 
the explanation of $f=5.4 r_{\rm{BLR}}^{0.3}$ \textbf{in Mrk 110 arising due to the radiation pressure from the accretion disc.} Thus, 
the relation of $f=6.8\mathscr{\dot M}_{f=1}^{0.4}$ reflects the physical essence of some relations similar to $f=5.4 r_{\rm{BLR}}^{0.3}$. 
The radiation pressure will produce more obvious \textbf{effects} on the BLR clouds as $\alpha$ increases. As the radiation pressure vanishes, 
$\alpha$ vanishes. The larger values of $f$ $\approx$ 8--16 make $M_{\rm{RM}}$ increase for Mrk 110. In like manner, \textbf{larger $f$ values}
may exist in quasars. Quasar J0100+2802 at $z = 6.30$, the most luminous quasar known at $z >6$, has $M_{\rm{RM}}(f=1)\sim$ $1.2\times 10^{10}M_{\rm{\odot}}$ 
and $L_{\rm{bol}}=1.62\times 10^{48}$ $\rm{ergs \/\ s^{-1}}$ \citep{Wu15}. It has $L_{\rm{Edd}}=1.5\times 10^{38}M_{\bullet}/M_{\odot}$
$\sim 1.8\times 10^{48}$ $\rm{ergs \/\ s^{-1}}$ for solar composition gas and $L_{\rm{bol}}/L_{\rm{Edd}}\sim$ 0.9. $\mathscr{\dot M}_{f=1}=$
$L_{\rm{bol}}/L_{\rm{Edd}}/\eta$ \citep{Du14}, where $\eta$ is the efficiency of converting rest-mass energy to radiation \citep{Th74}
and in general $\eta$ is \textbf{on} the order of 0.1. Thus $\mathscr{\dot M}_{f=1}\sim 9$ and $f\sim 16$ for J0100+2802. As $f\sim$ 16, 
$M_{\rm{RM}}\sim$ $1.9\times 10^{11}M_{\rm{\odot}}$ and $L_{\rm{bol}}/L_{\rm{Edd}}\sim$ 0.06. The larger black hole mass further gives rise 
to the most significant challenge to the Eddington limit growth of black holes in the early Universe \citep{Vo12,Wi10}. A high-$f$ value is
suggested for PG 1247+267 by the ultraviolet RM of carbon lines \citep{Tv14}. PG 1247+267 has $\lambda L_{\lambda}(1350\rm{\AA)}$ $=3.9 \times 10^{47}$ 
$\rm{ergs \/\ s^{-1}}$ and ionization stratification similar to low-luminosity AGNs. The broad H$\beta$ has a redward shift of 0.008 with respect to 
[O III] $\lambda 5007$ \citep{Mc99b}, and has $v_{\rm{FWHM}}=$ 7460 $\rm{km \/\ s^{-1}}$ \citep{Mc99a}. If $z_{\rm{g}}=0.008$, $f\approx$ 13, consistent 
with the high-$f$ value suggested in \citet{Tv14}. \textbf{Thus, the new method is capable of estimating} $f$ in quasars. \textbf{For the Eddington ratio
$L_{\rm{bol}}/L_{\rm{Edd}}(f)=L_{\rm{bol}}/fL_{\rm{Edd}}(f=1)=1$, we have $\mathscr{\dot M}_{f=1}=1133$, $f=113$ and  $\mathscr{\dot M}_{f}= 10$ 
in the case of $\eta=0.1$. Then, we have $L_{\rm{bol}}/L_{\rm{Edd}}=113$ in the case of assuming $f=1$. The fact that we don't often see sources 
with such (apparent) Eddington ratios may suggest that the Thomson cross-section typically used in the radiation pressure calculation is underestimating 
the coupling between the radiation and the BLR gas. If considering the line-driven radiation pressure \citep{Ca75}, the radiation pressure due to the gas
opacity will be $\sim 10^3$ times that due to the electron scattering opacity \citep{Fe09}.
}
\begin{figure}
\begin{center}
\includegraphics[angle=-90,scale=0.28]{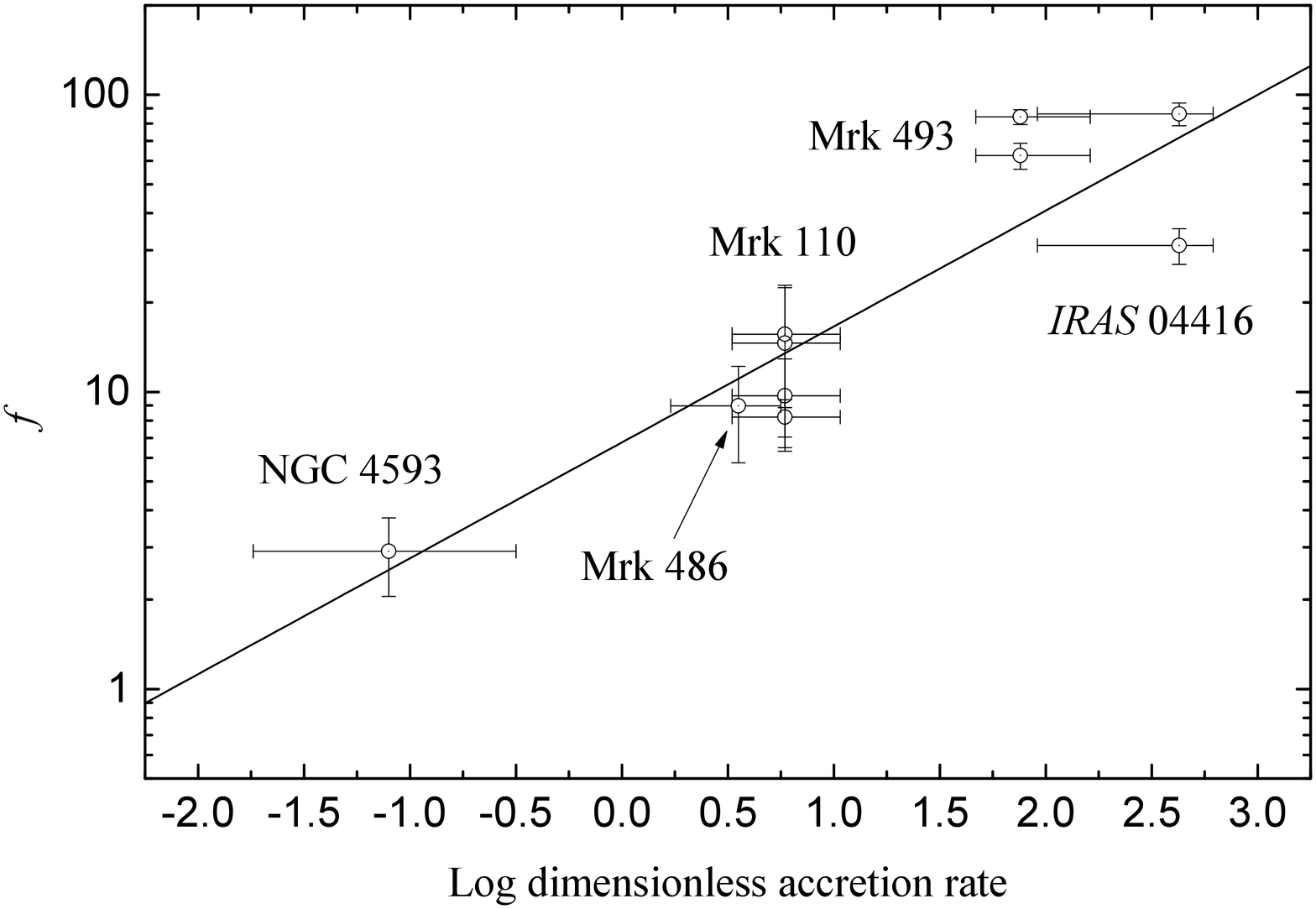}
\end{center}
 \caption{$f$ vs $\log \mathscr{\dot M}_{f=1}$. $\log \mathscr{\dot M}_{f=1}$ listed in Table 4. Black line is the best fit line to black circles
 with $r=0.901$ at the confidence level of 99.96 percent $[\log y=0.83(\pm 0.10)+0.39(\pm 0.07)\times x]$.}
  \label{fig3}
\end{figure}

With the large changes in the black hole mass estimates for these sources compared to the previous RM values, it would be interesting to 
know where they now fall on the $M_{\bullet}-\sigma_{\ast}$ relation. We find $\sigma_{\ast}$ measured by the stellar absorption lines for 
Mrk 110 and NGC 4593 in \textbf{the literature} (see Table 4). For \emph{IRAS} 04416, Mrk 493 and Mrk 486, the narrow line [O {\sc iii}] width 
$v_{\rm{FWHM}}$ can be converted into $\sigma_{\ast}$ by $\sigma_{\ast}=v_{\rm{FWHM}}$/2.35 \citep{Ne95} (see Table 4). The [O {\sc iii}] 
$v_{\rm{FWHM}}$ may serve as a good representation of $\sigma_{\ast}$ \citep{Ne00}, and \textbf{this method has been used in the literature} 
\citep[e.g.][]{Wa01}. $M_{\rm{grav}}$ and $\sigma_{\ast}$ are combined to show where they fall on the $M_{\bullet}-\sigma_{\ast}$ relation. 
Since the radiation pressure can significantly influence $f$, we use the $M_{\bullet}-\sigma_{\ast}$ relations derived for galaxies with $M_{\bullet}$ 
estimated by the gas, stellar and maser kinematics. \citet{Tr02} obtained for 31 nearby galaxies $\log M_{\bullet}/M_{\odot}=8.13+4.02\log$
$\sigma_{\ast}/\sigma_{0}$ with $\sigma_{0}=200$ $\rm{km \/\ s^{-1}}$. \citet{Wo13} obtained for 72 quiescent galaxies $\log M_{\bullet}/M_{\odot}$
$=8.37+5.31\log \sigma_{\ast}/\sigma_{0}$. The two relations are compared with the $M_{\rm{grav}}$ and $\sigma_{\ast}$ of five objects. 
\emph{IRAS} 04416, Mrk 486 and NGC 4593 follow the $M_{\bullet}-\sigma_{\ast}$ relations, and \textbf{Mrk 493 is consistent with either version 
of the relation at better than $2\sigma$} (see Figure 4). However, Mrk 110 lies significantly above the two relations (see Figure 4). \textbf{Mrk 110 
and Mrk 486 have nearly identical properties, with} $f$, $\mathscr{\dot M}_{f=1}$ and $M_{\rm{grav}}$ within the uncertainties. So, Mrk 110 and Mrk 486 
should have consistent $\sigma_{\ast}$. In fact, the $\sigma_{\ast}$ of Mrk 486 is basically two times as much as that of Mrk 110. \citet{Fe01} and 
\citet{Ne04} measured $\sigma_{\ast}$ for 6 and 16 Seyfert 1 galaxies by the Ca {\sc ii} triplet lines, respectively. \textbf{The spectra analysed 
by both papers have relatively low signal-to-noise ratios}. \textbf{The poor signal-to-noise and very} shallow absorption features in Mrk 110 might 
lead to smaller $\sigma_{\ast}$ \citep[see Figure 1 in][]{Fe01}. [O {\sc iii}] $\lambda 5007$ has $v_{\rm{FWHM}}=$ 336.4 $\rm{km\/\ s^{-1}}$ \citep{Do11}, 
and $v_{\rm{FWHM}}\approx$ 350 $\rm{km\/\ s^{-1}}$ found in the SDSS DR12\footnote {http://skyserver.sdss.org/dr12/en/tools/chart/navi.aspx}. Then 
[O {\sc iii}] $\lambda 5007$ has a mean $v_{\rm{FWHM}}\approx$ 343 $\rm{km\/\ s^{-1}}$ and $\sigma_{\ast}\approx 146$ $\rm{km\/\ s^{-1}}$ that makes 
Mrk 110 basically follow the $M_{\bullet}-\sigma_{\ast}$ relations in Figure 4. In the case of high-$f$ values given by the gravitational redshift 
method, these \textbf{five Seyfert galaxies are broadly consistent with} the $M_{\bullet}-\sigma_{\ast}$ relations in Figure 4. Thus, the higher $f$ 
values are reasonable, and the lower black hole masses estimated in the case of $\langle f\rangle \approx 1$ do not follow these relations.
\begin{figure}
\begin{center}
\includegraphics[angle=-90,scale=0.28]{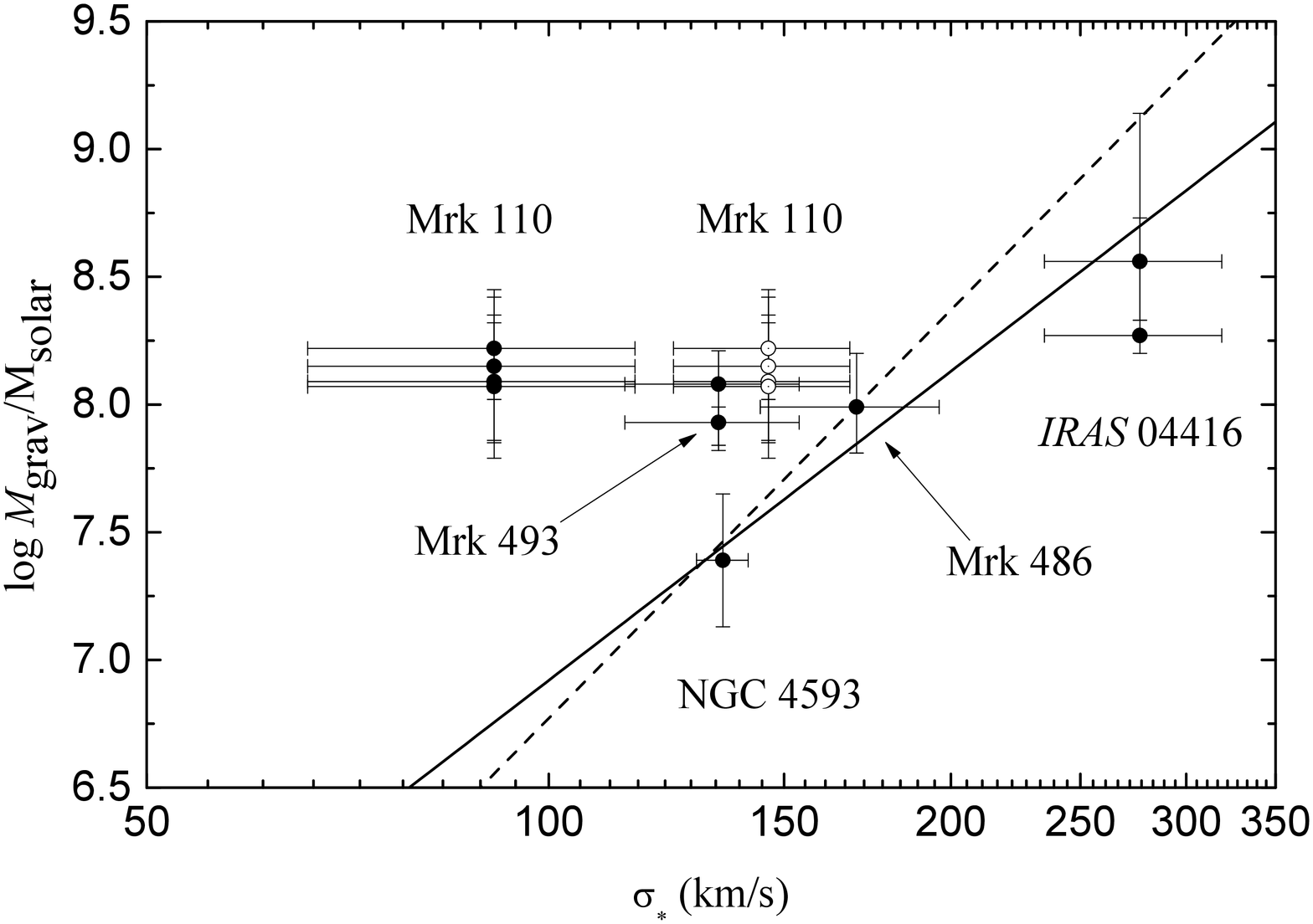}
\end{center}
 \caption{$\log M_{\rm{grav}}/M_{\rm{\odot}}$ vs $\sigma_{\ast}$. Solid line is the \citet{Tr02} relation for inactive galaxies.
 Dashed line is the \citet{Wo13} relation for quiescent galaxies. Open circles denote $\sigma_{\ast}$ converted from [O {\sc iii}]
 $v_{\rm{FWHM}}$ for Mrk 110.}
  \label{fig4}
\end{figure}
\begin{table}
\centering
\begin{minipage}{75mm}
\caption{$\mathscr{\dot M}_{f=1}$ and $\sigma_{\ast}$ for five Seyfert galaxies \label{tbl-4}}
\begin{tabular}{cccc}
\hline \hline

Object& $\log \mathscr{\dot M}_{f=1}$ & $\sigma_{\ast}$ $(\rm{km\/\ s^{-1}})$& Refs  \\
(1)&(2)&(3)&(4) \\

\hline

Mrk 110 & $0.77^{+0.26}_{-0.25}$ & $91\pm 25$$^\dag$&1  \\
        &                        & $146\pm 22$$^\ddag$&3  \\

NGC 4593 & $-1.10^{+0.60}_{-0.64}$ & $135\pm 6$$^\dag$&1  \\

\emph{IRAS} 04416 & $2.63^{+0.16}_{-0.67}$ & $277\pm 42$$^\ddag$&2  \\

Mrk 493 &  $1.88^{+0.33}_{-0.21}$ & $134\pm 20$$^\ddag$&2  \\

Mrk 486 & $0.55^{+0.20}_{-0.32}$ & $170\pm 26$$^\ddag$&2  \\

\hline

\end{tabular}
\\Notes: Column 1: object names; Column 2: logarithm of dimensionless accretion rate, taken from \citet{Du15}; Column 3:
the stellar velocity dispersion $\sigma_{\ast}$; Column 4: references of column (3). Refs 2 and 3 only give [O {\sc iii}] $v_{\rm{FWHM}}$. \\
\textbf{References}: (1) \citealt{Ne04}; (2) \citealt{Wa01}; (3) the SDSS DR12.
\\$^\dag$ $\sigma_{\ast}$ were measured by the stellar absorption lines.
\\$^\ddag$ $\sigma_{\ast}$ are converted from [O {\sc iii}] $v_{\rm{FWHM}}$, and the uncertainties are
taken to be 15\% of $\sigma_{\ast}$ \citep[e.g.][]{Fe01}.

\end{minipage}
\end{table}

\textbf{In some sources, blueshifted [O {\sc iii}] $\lambda 5007$ emission lines have been found} \citep[e.g.][]{Bo05,Ba14,Zh06}. 
The broad line shifts relative to [O {\sc iii}] $\lambda 5007$ \textbf{can thus acquire larger uncertainties} due to the [O {\sc iii}] 
blueshifts. \textbf{Given that uncertainty in the reference redshift, it would be beneficial to avoid using [O {\sc iii}] as the reference
for the broad line shifts.} Equation (7) only needs the redshift difference between two broad lines. Thus, the [O {\sc iii}] blueshifts 
will not influence $M_{\rm{grav}}$ given by equation (7). The consistent masses derived from equations (4) and (7) for Mrk 110 suggest 
that the [O {\sc iii}] blueshifts do not cause significant influence on $M_{\rm{grav}}$ given by equation (4) and the rms spectra. Thus, 
the broad-line-to-broad-line comparison path or the rms spectra could overcome the limitations of narrow line blueshifts. The mean spectra 
might overcome the same limitations. $z_{\rm{g}}$ can be given by the redward shift of the centroid of broad line with respect to its peak 
\citep{Jo16}. This approach \textbf{is another means of avoiding the limitations} of narrow line blueshifts. However, \textbf{cloud inflows 
could also generate a redward line shift, which could bias the gravitational redshift measurement.} \citet{Jo16} confirmed the gravitational 
redshift origin of this redward shift, \textbf{based on the theoretically expected relationship of $\Delta z_{\rm{50}}$ $\propto$ $v_{\rm{FWHM}}^2$}. 
The gravitational redshift is a natural outcome of the virialized motions of BLR clouds. The blueshifts are usual for high-ionization lines, 
e.g. broad C {\sc iv}, and are regarded as a signal of gas outflows \citep[e.g.][references therein]{Wa11}. The outflows may be driven by the 
radiation pressure of accretion disc. Mrk 486 has the redshifted broad Fe {\sc ii} and the blueshifted broad H$\beta$. These two shifted broad 
lines may be explained in terms of the combination of the outflows and the gravitational redshift effect. The redshifted broad H$\beta$ and 
Fe {\sc ii} in other 3 NLS1s might be dominated by the gravitational redshift.

\citet{Mu06} used the Kerr ray tracing simulations to study the gravitational redshifts of Mrk 110. When $r_{\rm{BLR}}$
$\gtrsim$ $100 r_{\rm{g}}$, the simulation results for stationarily rotating emitters are nearly identical to, within the errors,
those for static emitters in the Kerr space-time and the Schwarzschild space-time \citep[see Fig. 9 in][]{Mu06}. Mrk 110 has
$r_{\rm{BLR}}$ $\sim$ 560--4000 $r_{\rm{g}}$ for He II, He I, H$\beta$ and H$\alpha$. NGC 4593, Mrk 486, Mrk 493 and \emph{IRAS} 04416
have $r_{\rm{BLR}}$ $\sim$ 2700, 3100, 1700--2400 and 600--1300 $r_{\rm{g}}$, respectively. So, it is reasonable to estimate $f$ and
$M_{\rm{grav}}$ by the \textbf{use} of formulas in section 2. \emph{IRAS} 04416 and Mrk 493 have a mean $r_{\rm{BLR}}\sim 900 r_{\rm{g}}$
and $\sim 2000 r_{\rm{g}}$, respectively. According to \citet{Mu06}, the gravitational redshift can be probed out to $r_{\rm{BLR}}\sim$
$900 r_{\rm{g}}$ and $\sim 2000 r_{\rm{g}}$ with a resolution of $\approx$ 8.3 $\rm{\AA}$ and $\approx$ 3.8 $\rm{\AA}$, respectively.
The spectra measured in the RM campaign for SEAMBHs have a resolution of 1.8 $\rm{\AA}$ \citep{Du14,Hu15}. Thus, $z_{\rm{g}}$ can be
probed out to $r_{\rm{BLR}}\sim 900 r_{\rm{g}}$ and $\sim 2000 r_{\rm{g}}$, respectively, for \emph{IRAS} 04416 and Mrk 493 with the
spectra used to measure the redward shifts of H$\beta$ and Fe {\sc ii}. The spectral resolution of $\sim$ 500 $\rm{km \/\ s^{-1}}$
mentioned in \citet{Du14} and \citet{Hu15} is an instrumental broadening that mainly influences width of spectrum line, such as
$v_{\rm{FWHM}}$, but slightly influences the central wavelength of spectrum line. The instrumental broadening has been corrected
to obtain $v_{\rm{FWHM}}$ of broad lines in NLS1s \citep{Du14,Hu15}.

In this paper, based on the gravitationally redshifted optical broad emission lines in AGNs, a new method is proposed to measure
the virial factors $f$ in $M_{\rm{RM}}$ estimates by the \textbf{use} of $z_{\rm{g}}$ and $v_{\rm{FWHM}}$ of broad lines. First, this new
method is applied to NLS1 Mrk 110 with the gravitationally redshifted broad lines He II, He I, H$\beta$ and H$\alpha$. These four
lines have $f\approx$ 8--16 that are distinctly larger than $\langle f \rangle \approx$ 1. The He II and He I lines have slightly
smaller $f$ than the H$\beta$ and H$\alpha$ lines. There is a positive correlation between $f$ and $r_{\rm{BLR}}$ for Mrk 110,
$f=5.4 r_{\rm{BLR}}^{0.3}$ (see Figure 1), which can be naturally explained by the radiation pressure influence of accretion disc
on the BLR clouds. The radiation pressure seems to be more important than usually thought in AGNs. Second, NGC 4593 and Mrk 486 have
$f\approx 3$ and $f\approx 9$ given by the redward shifted H$\alpha$ and Fe {\sc ii} lines, respectively. Third, NLS1s Mrk 493,
\emph{IRAS} 04416 and Mrk 42 have high-$f$ values of several tens given by the redward shifted H$\beta$ and Fe {\sc ii} lines. In
the plot of $\log f$ versus $\log r_{\rm{BLR}}$, NGC 4593 is below the best fit line to Mrk 110, Mrk 486 is around the best fit line,
and both of Mrk 493 and \emph{IRAS} 04416 are above the same line (see Figure 1). These differences result from the accretion rate
differences. NGC 4593, Mrk 486, Mrk 110, Mrk 493 and \emph{IRAS} 04416 have $\mathscr{\dot M}_{f=1}=$ 0.08, 3.55, 5.89, 75.9 and 426.6 or
$\mathscr{\dot M}_{\rm{f}}=$ 0.03, 0.40, 0.49, 1.04 and 7.27, respectively. \textbf{We find that} $f=6.8\mathscr{\dot M}_{f=1}^{0.4}$ for 
these five objects \textbf{(see Figure 3), which could be due to radiation pressure.} The black hole masses $M_{\rm{grav}}$ match
the $M_{\bullet}-\sigma_{\ast}$ relations derived from galaxies with $M_{\bullet}$ estimated by the gas, stellar and maser kinematics
(see Figure 4). The measurements of $f$ by $z_{\rm{g}}$ of broad lines have the important potential to improve the researches
on AGNs. The larger $f$ values given by $z_{\rm{g}}$ \textbf{can generate higher black hole masses and lower Eddington ratios.}

\section*{Acknowledgments}
We are grateful to the anonymous referee for important comments leading to significant improvement of this paper.
We thank the helpful discussions of Dr. H. Q. Li, Dr. P. Du and Dr. F. Wang. HTL thanks the National Natural Science
Foundation of China (NSFC; grants 11273052 and U1431228) for financial support. JMB acknowledges the support of the
NSFC (grant 11133006). HTL also thanks the financial supports of the project of the Training Programme for the Talents
of West Light Foundation, CAS and the Youth Innovation Promotion Association, CAS.

\label{lastpage}

\end{document}